\documentclass{article}

\usepackage{microtype}
\usepackage{graphicx}
\usepackage{caption}
\usepackage{subcaption}
\usepackage{booktabs} 
\usepackage{xspace}

\usepackage{hyperref}

\usepackage[normalem]{ulem}

\usepackage[accepted]{icml2024}
\usepackage{algorithm}
\usepackage{algorithmic}

\usepackage{amsmath}
\usepackage{amssymb}
\usepackage{mathtools}
\usepackage{amsthm}
\usepackage{amsfonts}

\usepackage[capitalize,noabbrev]{cleveref}

\theoremstyle{plain}

\theoremstyle{definition}

\theoremstyle{remark}

\DeclareMathOperator*{\argmax}{arg\,max}

\newcommand{\Sys}{MuxServe\xspace}

\begin{document}

\date{}

\twocolumn[
\icmltitle{\Sys: Flexible Spatial-Temporal Multiplexing for Multiple LLM Serving}



\icmlsetsymbol{equal}{*}

\begin{icmlauthorlist}
\icmlauthor{Jiangfei Duan}{cuhk,shlab}
\icmlauthor{Runyu Lu}{hust}
\icmlauthor{Haojie Duanmu}{shlab,sjtu}
\icmlauthor{Xiuhong Li}{pku}
\icmlauthor{Xingcheng Zhang}{shlab}
\icmlauthor{Dahua Lin}{cuhk,shlab}\\
\icmlauthor{Ion Stoica}{ucb}
\icmlauthor{Hao Zhang}{ucsd}
\end{icmlauthorlist}

\icmlaffiliation{cuhk}{The Chinese University of Hong Kong}
\icmlaffiliation{shlab}{Shanghai AI Laboratory}
\icmlaffiliation{pku}{Peking University}
\icmlaffiliation{ucsd}{University of California San Diego}
\icmlaffiliation{ucb}{UC Berkeley}
\icmlaffiliation{hust}{Huazhong University of Science and Technology}
\icmlaffiliation{sjtu}{Shanghai Jiao Tong University}

\icmlcorrespondingauthor{Hao Zhang}{haozhang@ucsd.edu}
\icmlcorrespondingauthor{Xiuhong Li}{lixiuhong@pku.edu.cn}

\icmlkeywords{Serving, Spatial-Temporal Multiplexing, Large Language Model}

\vskip 0.3in
]


\printAffiliationsAndNotice{}

\begin{abstract}
Large language models (LLMs) have demonstrated remarkable performance, and organizations are racing to serve LLMs of varying sizes as endpoints for use-cases like chat, programming and search. However, efficiently serving multiple LLMs poses significant challenges for existing approaches due to varying popularity of LLMs.
In the paper, we present \Sys, a flexible spatial-temporal multiplexing system for efficient multiple LLM serving. The key insight behind is to colocate LLMs considering their popularity to multiplex memory resources, and leverage the characteristics of prefill and decoding phases to separate and flexibly colocate them to multiplex computation resources. 
\Sys formally formulates the multiplexing problem, and proposes a novel placement algorithm and adaptive batch scheduling strategy to identify optimal colocations and maximize utilization. 
\Sys designs a unified resource manager to enable flexible and efficient multiplexing. Evaluation results show that \Sys can achieves up to $1.8\times$ higher throughput or processes $2.9\times$ more requests within $99\%$ SLO attainment. The code is available at: \url{https://github.com/hao-ai-lab/MuxServe}.
\end{abstract}



\section{Introduction}

Recent advances in large language models (LLMs) are transforming the AI industry \cite{gpt-3, foundation-model-survey, palm}. A variety of versions and scales of LLMs have been pretrained and finetuned for various use cases, such as chat, programming, and search. Many organizations, such as Google, OpenAI, Huggingface, are racing to serve these LLMs as endpoints to their users. However, the unprecedented capabilities of LLMs come at a significant inference cost -- serving a single 175B LLM~\cite{gpt-3}  requires eight A100 (80GB) GPUs;
efficiently serving \emph{multiple LLMs}, each catering to different group of users and needs, are even costlier and have emerged as a crucial and time-sensitive demand within the community, especially for LLM endpoint providers.

\begin{figure}[t]
    \centering
    \includegraphics[width=\linewidth]{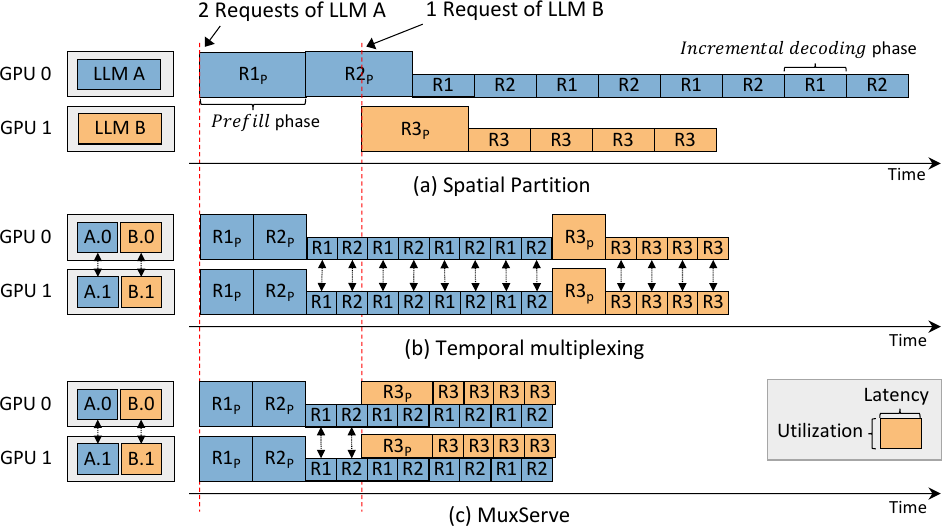}
  \caption{Three multiplexing strategies and GPU utilization of serving two LLMs on two GPUs.}
  \label{fig:intro}
  \vskip -0.2in
\end{figure}
\begin{figure}[t]
    \centering
    \includegraphics[width=\linewidth]{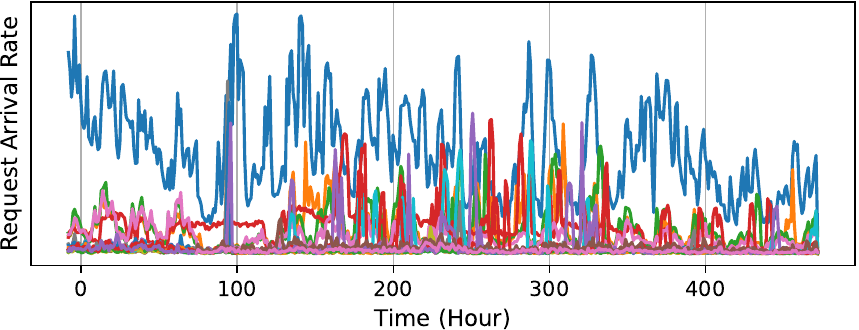}
    \caption{The dynamic request arrival rates of different LLMs over a 20 day period.}
    \label{fig:traffic}
    \vskip -0.1in
\end{figure}

To serve multiple LLMs with a cluster of resources, existing systems~\cite{TGI, triton-server, vllm} typically use \emph{spatial partitioning} (\Cref{fig:intro}a), which involves allocating separate groups of GPUs for each LLM to accommodate their large model size and the key-value cache (KV cache) generated during inference. However, this spatial partition approach often leads to significant under-utilization of GPUs. Figure~\ref{fig:traffic} shows real traffic observed by an LLM endpoint provider in 20 days: Different LLMs typically exhibit varying levels of popularity among users influenced by factors such as output quality, response speed, and usage patterns. 
Spatial partitioning disregards the varying popularity of different LLMs -- LLMs with low arrival rates tend to receive sparse requests, resulting in idle GPUs for extended periods (as illustrated by GPU 1 in \Cref{fig:intro}a). Conversely, popular LLMs experience a substantial burden in handling incoming requests (GPU 0 in \Cref{fig:intro}a), leading to a potential performance bottleneck.

Another line of work explores temporal multiplexing (\Cref{fig:intro}b) to serve multiple large models~\cite{alpaserve}, resulting in reduced serving latency in the presence of bursty workloads. This approach involves partitioning models onto a shared group of GPUs using intra- and inter-operator parallelism, and scheduling requests in an interleaved manner to share the computation and memory resources. However, this approach does not fully leverage the potential of GPUs when serving multiple LLMs, as it overlooks the unique characteristics of the \textit{prefill} and \textit{incremental decoding} phases of autoregressive LLMs. The \textit{incremental decoding} phase, which typically plays a significant role in the inference process, often falls short in fully utilizing GPUs. Therefore, temporal multiplexing brings a wave-shaped utilization change, and most of the time it is in the trough (\Cref{fig:intro}b).




In this work, we explore to serve multiple LLMs with flexible \emph{spatial-temporal multiplexing} to improve GPU utilization (\Cref{fig:intro}c) motivated by the following two key insights. 
Firstly, since \textit{prefill} and \textit{incremental decoding} phases have distinct computation characteristics, we separate them into different jobs and flexibly colocate prefill or decoding jobs from \emph{different LLMs} to multiplex computation resources. 
Secondly, we colocate LLMs considering their popularity to multiplex memory resources and improve utilization. In \Cref{fig:intro}c, request of LLM B can be scheduled at its arrival since the \textit{incremental decoding} phase of LLM A cannot fully utilize the GPUs. This flexible multiplexing allows \Sys to finish all the requests in a shorter time, thus improving utilization.

We design and implement \Sys to enable flexible and efficient spatial-temporal multiplexing for multiple LLM serving. Given a cluster configuration, a set of LLMs with workloads, \Sys first formulates an optimization problem to search for the optimal colocations and batching and scheduling strategy (\Cref{method:formulation}). To efficiently solve the problem, \Sys proposes an enumeration-based greedy placement algorithm (\Cref{method:place}) and adaptive batch scheduling algorithm (\Cref{method:fairness}) to maximize utilization while ensuring fair sharing among LLMs. 
We discover that spatial-temporal partitioning is achieved by partitioning GPU SMs using CUDA MPS~\cite{mps}, and \Sys designs a novel unified resource manager (\Cref{method:res_manager}) to enable efficient multiplexing. We finally evaluate \Sys with both synthetic and real workload on a 32-GPU cluster. Evaluation results show that \Sys achieves up to $1.8\times$ higher throughput compared to prior state-of-the-art systems.


In summary, \Sys makes the following contributions,
\begin{enumerate}
    \item The first to explore spatial-temporal multiplexing for LLM serving and formally formulate the problem.
    \item A novel placement algorithm and adaptive batch scheduling strategy to determine the best collocations and maximize utilization.
    \item A viable system design and implementation with unified resource manager to enable efficient multiplexing of LLMs and comprehensive evaluation. 
\end{enumerate}

\section{Background and Motivation}
\subsection{LLM Inference}
\label{bg:llm_infer}
LLMs stack Transformer~\cite{transformer} blocks and each block consists of multi-head attention and feed-forward networks. Given input prompts, LLMs generate output tokens in autoregressive manner. The inference process includes two phases: \textit{prefill} and \textit{incremental decoding}. In \textit{prefill} phase, LLMs process the entire prompt tokens in parallel and generate the first output token. Subsequently, in \textit{decoding} phase, LLMs iteratively generate one output token at a time, building upon previously generated token. During inference, LLMs save key-value cache (KV cache) for each token, which dynamically increases as tokens are produced.

The two phases of LLM inference exhibit dinstict characteristics. The \textit{prefill} phase, characterized by long input prompts, heavily utilizes computation resources, while the \textit{incremental decoding} phase, with limited generated tokens, results in insufficient GPU utilization despite dominating the inference process due to the need to generate lengthy outputs for each prompt. For example, the average prompt and output length is 161 and 338 tokens in ShareGPT~\cite{sharegpt}, respectively.


\begin{figure}[t]
    \centering
    \includegraphics[width=0.98\linewidth]{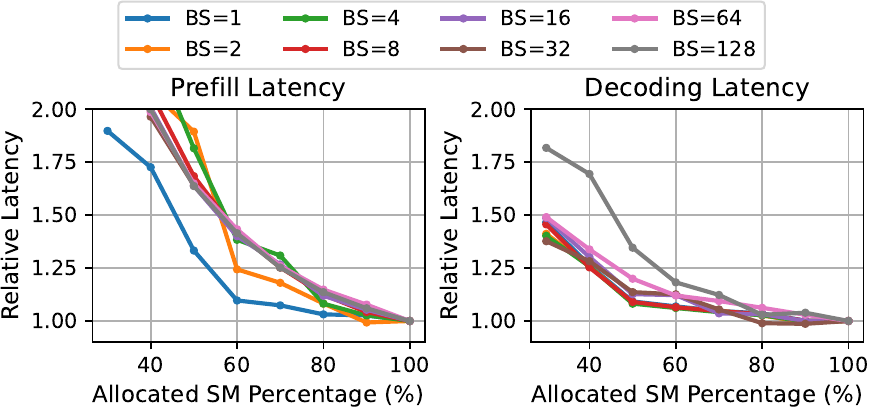}
  \caption{Relative batch inference latency as the fraction of computing resources assigned to LLaMA-7B changes from $30\%$ to $100\%$. The input sequence length is 128.}
  \label{fig:sm_eff}
  \vskip -0.1in
\end{figure}

\subsection{Distributed LLM Inference}
Distributed inference is introduced to accommodate LLMs that cannot fit in a single GPU or accelerate inference process, and includes two categories.
\textit{Intra-operator parallelism}~\cite{alpa} splits individual LLM layers across multiple GPUs and requires collective communications to transform input and output tensors during inference. It can significantly reduce inference latency, but also introduces additional communication overheads.
\textit{Inter-operator parallelism} partitions LLMs into multiple stages. Each stage is placed on one GPU with data dependency, and executed in a pipeline fashion (i.e. pipeline parallelism~\cite{pipedream}). Pipeline parallelism comes with negligible overhead, but does not reduce inference latency.

\subsection{LLM Popularity Varies}
\Cref{fig:traffic} displays the serving traffic of multiple LLMs over 20 days, as observed from an LLM endpoint provider.
It is evident that the popularity varies significantly, and each LLM experiences distinct and changing arrival rates. Popular LLMs (blue line) consistently receive a considerably higher volume of serving traffic compared to other LLMs, resulting in higher resources demands. In contrast, less popular LLMs may exhibit consistently low arrival rates throughout the observed period, occupying fewer resources. This dynamic and diverse nature of request arrival rates emphasizes the need for a flexible and adaptive approach to efficiently serve multiple LLMs based on their individual popularity and demand, which would translate into significant cost reduction for LLM endpoint providers.

\subsection{Multiplexing Opportunity on GPU}
\label{bg:opportunity}
\emph{Spatial multiplexing} is a resource sharing technique that splits the GPU resource (memory or/and SMs, i.e. Streaming Multiprocessors) into smaller partitions. Each partition is then allocated to perform different tasks simultaneously. \emph{Temporal multiplexing} enables sharing of GPUs where each task occupies the entire computation resource during a specific time interval. With multiplexing, GPUs can handle multiple workloads concurrently, leading to increased efficiency and throughput. NVIDIA also offers Multi-Instance GPU (MIG)~\cite{mig} to split memory and SMs into independent instances, and CUDA MPS~\cite{mps} to partition SMs for different processes.

Prior works~\cite{gslice, mig-eval} have explored spatial multiplexing to serve multiple DNNs by assigning one DNN to a separate partition for enhanced performance. However, serving LLMs presents non-trivial challenges due to their unique characteristics. A significant difference lies in the memory bottleneck: \textit{the huge memory requirements render previous approaches ineffective since it is unfeasible to hold multiple LLMs in a single GPU.}

To mitigate the memory bottleneck, AlpaServe~\cite{alpaserve} involves parallelism to distribute several large models on multiple GPUs and utilizes temporal multiplexing to serve these models. However, temporal multiplexing ignores the characteristics of prefill and decoding phases of LLMs.
As illustrated in \Cref{fig:sm_eff}, when the amount of computation resources allocated to the dominant \textit{decoding} phase is reduced, it does not lead to a substantial decrease in latency or throughput. Moreover, parallelization across multiple GPUs further reduces the computation requirements of LLMs. Temporal multiplexing thus results in significant resource under-utilization.

Recognizing the distinct resource requirements of \textit{prefill} and \textit{decoding} phases, we reveal that different LLMs can be colocated to multiplex the computation resources flexibly for improved efficiency and utilization. In the following sections, we will present \Sys, the first system that explores spatial-temporal multiplexing for multiple LLM serving.


\section{Method}

\subsection{Problem Formulation}
\label{method:formulation}
Consider we have a cluster $C$ and a set of LLMs $M$ with workload $W$\footnote{Suppose the workload is known. Otherwise, the workload can be estimated from history traffic since it changes slowly.} to be served, one key insight \Sys leveraged to improve system utilization is to colocate LLMs considering their popularity. LLMs with high request rates can be colocated with LLMs with low request rates to efficiently utilize resources. To achieve this, we introduce \textit{LLM unit}, which refers to a group of LLMs that will be colocated together with the GPUs they are assigned. Our goal is to find the best group of \textit{LLM units} $B^*$ that maximize GPU utilization (i.e. throughput), hence the problem can be formulated as,

\begin{equation}
\begin{aligned}
\label{eq:inter_unit}
   B^* = \argmax_{B\in \mathcal{B}} \sum_{b\in B} \texttt{F}(b, W_{b})
\end{aligned}
\end{equation}

where $\mathcal{B}$ represents all possible LLM units group, and \texttt{F}($\cdot$, $\cdot$) estimates the throughput for a unit $b$ with workload $W_{b}$. 

Within an LLM unit, \Sys leverages the insight that \textit{prefill} and \textit{decoding} phases of LLM inference exhibit distinct computation resources requirements, and splits them into \textit{prefill} and \textit{decoding} jobs. Each job occupies a fixed amount of SM resources and executes a prefill or decoding step for a batch requests of an LLM. Different jobs can be flexibly colocated to share the computation and memory resources. 
However, as there are multiple LLMs with distinct workload characteristics, different batching and scheduling strategies can lead to different throughputs, and different LLMs may also compete for resources. Therefore, given an LLM unit $b$ that contains colocated LLMs $b_{llm}$, we need to find the optimal batching and scheduling strategy $S$ that can maximize the throughput of the entire unit ,while ensuring fair resource sharing among LLMs within the unit. Therefore, the problem $\texttt{F}(b, W_b)$ can be formulated as,

\begin{equation}
\begin{aligned}
\label{eq:intra_unit}
    \texttt{F}(b, W_b) = \max_{S} \sum_{m \in b_{llm}} \texttt{tpt}_S(m, b, W_b) \quad s.t. \\
    |\texttt{R}(m_i, W_{m_i}) - \texttt{R}(m_j, W_{m_j})| \le \epsilon, \forall m_i, m_j \in b_{llm} 
\end{aligned}
\end{equation}

where $\texttt{tpt}_S$($\cdot$, $\cdot$, $\cdot$) estimates the throughput of an LLM $m$ in the unit $b$ using strategy $S$, \texttt{R}($\cdot$, $\cdot$) estimates the normalized computation or memory resources consumption of an LLM $m$ with workload $W_m$, and $\epsilon$ is a small number ensuring fairness. 
\texttt{R}($\cdot$, $\cdot$) is normalized to account for varying LLM scales and popularity, since large and popular LLMs typically requires more resources.


\begin{algorithm}[t]
   \caption{Enumeration-based Greedy LLM Placement}
   \label{alg:placement}
\begin{algorithmic}
   \STATE {\bfseries Input:} LLM list $M$, cluster $C$, workload $W$
   \STATE {\bfseries Output:} The optimal group of LLM unit $best\_units$
   \STATE $\hat{M} \leftarrow$ llm\_parallel\_candidate($M$, $W$) \textcolor{gray}{// \Cref{alg:candidate}}
   \STATE $\mathcal{D} \leftarrow$ get\_potential\_device\_mesh\_groups($C$, $M$)

   \STATE $best\_tpt, best\_units \leftarrow 0, None$
   \FOR{$D \in \mathcal{D}$}
   \STATE \textcolor{gray}{// Greedily place LLMs on mesh group $D$}
   \STATE $M'$=sort($\hat{M}$, $key$=$computation$, $descend$=$True$)
   \FOR{$m \in M'$}
   \STATE $best\_mesh, max\_delta$ $\leftarrow$ None, -1
   \FOR{$d \in D$}
   \STATE $u$ = make\_unit($m$, $d$)
   \STATE $delta$ = \texttt{F}($u$, $W_u$) - \texttt{F}($d.u$, $W_{d.u}$)
   \IF{$delta > max\_delta$}
   \STATE $best\_mesh, max\_delta$ = $d, delta$
   \ENDIF
   \ENDFOR
   \STATE $best\_mesh.u$ = make\_unit($m$, $best\_mesh$)
   \ENDFOR

   \STATE $tpt$ = sum(\texttt{F}($d.u$, $W_{d.u}$) for $d\in D$)
   \IF{$best\_tpt < tpt$}
   \STATE $best\_tpt, best\_units \leftarrow tpt$, [$d.u$ for $d\in D$]
   \ENDIF
   \ENDFOR
   \vskip -0.2in
\end{algorithmic}
\end{algorithm}

Given the formulation above, we first introduce our placement algorithm to solve the problem (\Cref{eq:inter_unit}) in \Cref{method:place}, which will maximize the intra-unit throughput (\Cref{eq:intra_unit}) with our batching and scheduling strategy (\Cref{method:fairness}). Finally we describe our unified resource manager to enable efficient multiplexing in \Cref{method:res_manager}.

\begin{algorithm}[t]
   \caption{LLM Parallel Candidate Generation}
   \label{alg:candidate}
\begin{algorithmic}
   \STATE {\bfseries Input:} LLM list $M$, workload $W$
   \STATE {\bfseries Output:} The parallel candidate $\hat{M}$
   \STATE $\hat{M} \leftarrow []$
   \FOR{$m \in M$}
   \STATE $sm\_list \leftarrow$ get\_potential\_sm\_list($m$)
   \STATE $tp\_list \leftarrow$ get\_potential\_tp\_degree($m$)
   \FOR{$p \in tp\_list$}
   \FOR{$num\_sm \in$ sorted($sm\_list$)}
   \STATE $tpt, bs \leftarrow$ estimate\_throughput($m, num\_sm, p$)
   \IF{$tpt \ge W_m$}
   \STATE $m.candidate$.add(($p, num\_sm, bs$))
   \STATE {\bfseries break}
   \ENDIF
   \ENDFOR
   \ENDFOR
   \STATE $\hat{M}$.append($m.candidate$)
   \ENDFOR
\end{algorithmic}
\end{algorithm}

\subsection{Placement Algorithm}
\label{method:place}

Determining the optimal group of LLM units poses a challenging combinatorial optimization problem. As the number of devices and LLMs increases, the total number of possible LLM unit combinations grows exponentially. To solve \Cref{eq:inter_unit} efficiently, we design an enumeration-based greedy algorithm as outlined in \Cref{alg:placement}. 
The insight behind is to prioritize the placement selection for LLMs with large computation requirements, which considers both the model scale and popularity. With this algorithm, \Sys can find a good solution efficiently.

In \Cref{alg:placement}, \Sys first calls \Cref{alg:candidate} to generate all possible parallel candidates $\hat{M}$ considering the workload $W$. A parallel candidate refers to a configuration that meets the workload requirements while utilizing the fewest number of SMs. 
For each LLM $m$, \Sys enumerates all possible combinations of configurations by varying the number of SMs and intra-operator parallelism degrees to find a set parallel candidates. 
For each intra-operator parallelism degree, \Sys has one possible parallel candidate.

\Sys then enumerates all potential device mesh groups to find the best LLM units. Each mesh comprises several GPUs that will be used to serve a set of LLMs concurrently. Given a device mesh group $D$ and a list of parallel partition candidates $\hat{M}$, \Sys greedily places LLMs on meshes to find the optimal group of LLM units. \Sys prioritizes the placement selection for LLMs with large computation requirements to maximize serving utilization.
For a specified LLM $m$, \Sys iterates over all available meshes and approximates the expected increase in throughput with \texttt{F}($\cdot$, $\cdot$). The LLM $m$ is then placed on the mesh that yields the maximum throughput increase. This process is repeated for all LLMs. Subsequently, \Sys estimates the serving throughput of the LLM units after the placement, and selects the LLM units that offer the best serving throughput as the optimal group of LLM units.

The complexity of \Cref{alg:placement} is $O(MCD)$, where $M$ is the number of LLMs, $C$ is the number of devices and $D$ is the number of potential device mesh groups. Given a large cluster, enumerating all possible device mesh groups can be slow. We prune the search space effectively incorporating the following heuristics: the intra-operator parallelism is typically adopted within a node, and workload imposes constraints on the possible mesh size.

\begin{algorithm}[t]
   \caption{Adaptive Batch Scheduling (ADBS)}
   \label{alg:adbs}
\begin{algorithmic}
   \STATE {\bfseries Input:} LLM list $M$
   \STATE $prefill\_waiting$ $\leftarrow$ false
   \STATE $quota \leftarrow$ init\_token\_block\_quota($M$)
   \WHILE{True}
   \IF{no prefill jobs in execution}
   \STATE $prefill\_waiting \leftarrow$ True
   \STATE $m \leftarrow$ round-robin a prefill job from $M$
   \IF{resource\_enough($m$, $quota$)}
   \STATE execute\_and\_update($m$, $quota$)
   \STATE $prefill\_waiting \leftarrow$ False
   \ENDIF
   \ENDIF

    \IF{not $prefill\_waiting$}
    \STATE $m \leftarrow$ round-robin a decoding job from $M$
    \WHILE{resource\_enough($m$, $quota$)}
    \STATE execute\_and\_update($m$, $quota$)
    \STATE $m \leftarrow$ round-robin a decoding job from $M$
    \ENDWHILE
    \ENDIF

    $quota =$ adapt\_quota\_periodically($M$, $quota$)
    \ENDWHILE
\end{algorithmic}
\end{algorithm}

\subsection{Maximize Intra-unit Throughput}
\label{method:fairness}

If there is only one LLM in a unit, the situation is reduced to single LLM serving, which has been extensively studied. However, when it comes to multiplexing multiple colocated LLMs with dynamically varying requests arrival times (\Cref{eq:intra_unit}), the solution is non-trivial due to the following challenges: requests for different LLMs cannot be batched together, LLMs in the unit have distinct workload characteristics, and different LLMs may compete for resources. It is impractical to find an optimal exact solution due to the complexity of the problem.

To address these challenges, we first define \texttt{R}($\cdot$, $\cdot$) as the token block usage~\citep{llmfairness} of an LLM, based on the observation that KV cache size poses a significant performance bottleneck for LLM serving. \Sys assigns a token block quota to each LLM to ensure fair sharing. Counting token blocks provides a more intuitive way to consider the scale of LLMs, as tokens from different LLMs consume varying amounts of KV cache. Moreover, to consider variations in workload characteristics, \texttt{R}($\cdot$, $\cdot$) is also normalized by request rates.

Then we maximize the intra-unit throughput by exploring \textit{prefill-decoding} and \textit{decoding-decoding} collocation. \Sys prioritizes prefill jobs and fills remaining resources with decoding jobs. This is motivated by the observation that \textit{decoding} jobs of a single LLM typically requires a few computation resources and can be batched together. Prioritizing prefill jobs can maximize the opportunity for colocation and batching. 





\Cref{alg:adbs} describes our \textit{adaptive batch scheduling} (ADBS) algorithm to maximize intra-unit throughput while ensuring fairness. ADBS first restricts the usage of token blocks by setting a quota for each LLM. 
In the main loop, if there is no prefill jobs in execution, ADBS employs a round-robin approach to select and execute a prefill job from served LLMs. If the resource is not enough, ADBS stops scheduling decoding jobs until resource capacity is met. Otherwise, ADBS schedules decoding jobs with round-robin until resource is not enough to maximize colocation.


To further improve utilization, ADBS adapts the token block quota for each LLM periodically. During runtime, \Sys monitors the KV cache utilization. \Sys identifies low-utilization LLMs and proactively transfers KV cache blocks from these LLMs to high-utilization LLMs. This dynamic reallocation of KV cache blocks ensures optimal utilization of resources and promotes efficient sharing among the LLMs within a unit.

ADBS approximates the solution of \Cref{eq:intra_unit} to maximize intra-unit throughput. But the concrete throughput $\texttt{tpt}_S(\cdot, \cdot, \cdot)$ cannot be obtained without profiling. To address this, we build a simple yet effective analytical estimator to estimate the throughput of LLM $m$ with


\begin{equation}
    \label{eq:tpt}
    \texttt{tpt}_S(m, b, W_b) = min\{\frac{b^{m}}{\sum_{i\in b}t^{i}_{p} + t^{m}_{d} \cdot l^{m}_{o}}, W_b\}
\end{equation}

where $t^{m}_{p}, t^{m}_{d}$ and $l^{m}_{o}$ represent the prefill latency, decoding latency and average generation length of a batch requests with size $b^{m}$ for LLM $m$, respectively. This formulation is based on the observation that prefill phases of different LLMs are generally executed sequentially and decoding phases can be executed concurrently, and different phases are interleaved. Therefore, the latency of a batch requests is equal to the sum of all prefill phases of different LLMs and the decoding phases of the LLM.
The prefill and decoding latency of different batches and request length can be profiled in advance. The average generation length can be estimated from requests history or specific dataset, ShareGPT~\cite{sharegpt} for instance. 
Given the request arrival rates, we can use binary search to find the batch size $b$ that can satisfy the traffic. More details about the formulation please refers to \Cref{app:cost_model}.

\subsection{Resource Manager}
\label{method:res_manager}

\begin{figure}[t]
    \centering
    \includegraphics[width=\linewidth]{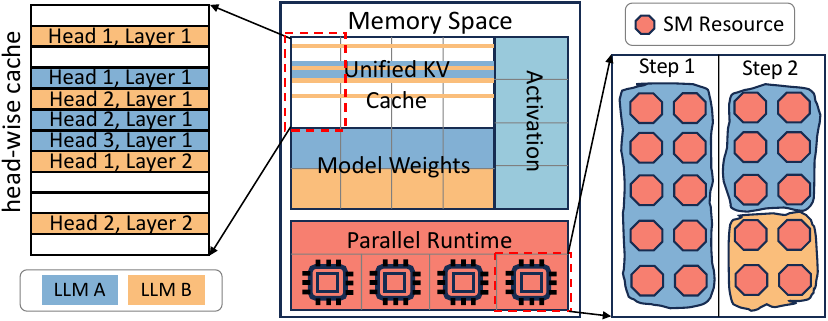}
  \caption{Overview of GPU resource management in an LLM unit. The memory is divided into 3 partitions to store KV cache, weights and activations, respectively. The parallel runtime partitions SM dynamically to different jobs.
  }
  \label{fig:arch}
  \vskip -0.1in
\end{figure}

After finding the optimal LLM units and determining the batching and scheduling strategy, \Sys requires a new mechanism to support flexible and efficient spatial-temporal multiplexing of LLMs due to the following challenges: different prefill and decoding jobs need to flexibly share the computation resources, and share the weights and KV cache to reduce memory waste and fragmentation. To address these challenges, \Sys proposes a unified resource manager for flexible and efficient multiplexing. Each LLM unit hosts a unified resource manager, and \Cref{fig:arch} shows the overview of GPU resource management in an LLM unit.

The parallel runtime manages computation resources of an LLM unit in the granularity of SM based on NVIDIA MPS~\cite{mps}. \Sys schedules prefill and decoding jobs from colocated LLMs with ADBS algorithm, then the parallel runtime dynamically assigns SMs to each job at runtime rather than statically allocating. This enables \Sys to flexibly multiplex the execution of different LLMs. As illustrated in the right part of \Cref{fig:arch}, the SMs are all allocated to one job at step 1 due to its large computation intensity. After the execution, \Sys schedules two jobs that can be executed concurrently at step 2 to share the SM resources.


The prominent challenge lies in sharing the memory resources among different jobs is to reduce memory waste and fragmentation. LLM weights and KV cache consume a huge amount of memory and need to be shared among jobs. Furthermore, KV cache increases dynamically, and different LLMs posses varying sizes of KV cache due to differences in the number of attention heads, hidden layers, and hidden sizes. 


To efficiently share memory resources, \Sys divides them into three partitions. The first partition is a unified KV cache space enabled by our head-wise cache. Leveraging the observation that the size of each attention head is often consistent or has limited variations across different LLMs, for example LLaMAs~\cite{llama} and GPT-3~\cite{gpt-3} all use 128. \Sys divides the KV cache table into small blocks, and each block holds the KV cache of one head for several tokens. This head-wise granularity enables \Sys to accommodate the KV cache of different LLMs in a unified space to share the memory.
To reduce redundancy, the second partition stores a single replica of LLM weights that can be shared among prefill and decoding jobs. The final partition reserves space for activation, which is utilized during inference runtime. 

\Sys adopts a unified KV cache instead of reserving separate KV cache for each LLM. This shared cache enables \Sys to dynamically adjust the cache allocation during runtime with minimal overhead. As a result, \Sys can handle  bursty and changing workload better.
Notably, vLLM~\cite{vllm} proposes Paged Attention to improve memory utilization for single LLM serving, while our unified KV cache addresses a distinct scenario, where multiple LLMs of varying sizes, popularities, and configurations need to share the cache. 

\begin{figure*}[h]
    \centering
    \includegraphics[width=\linewidth]{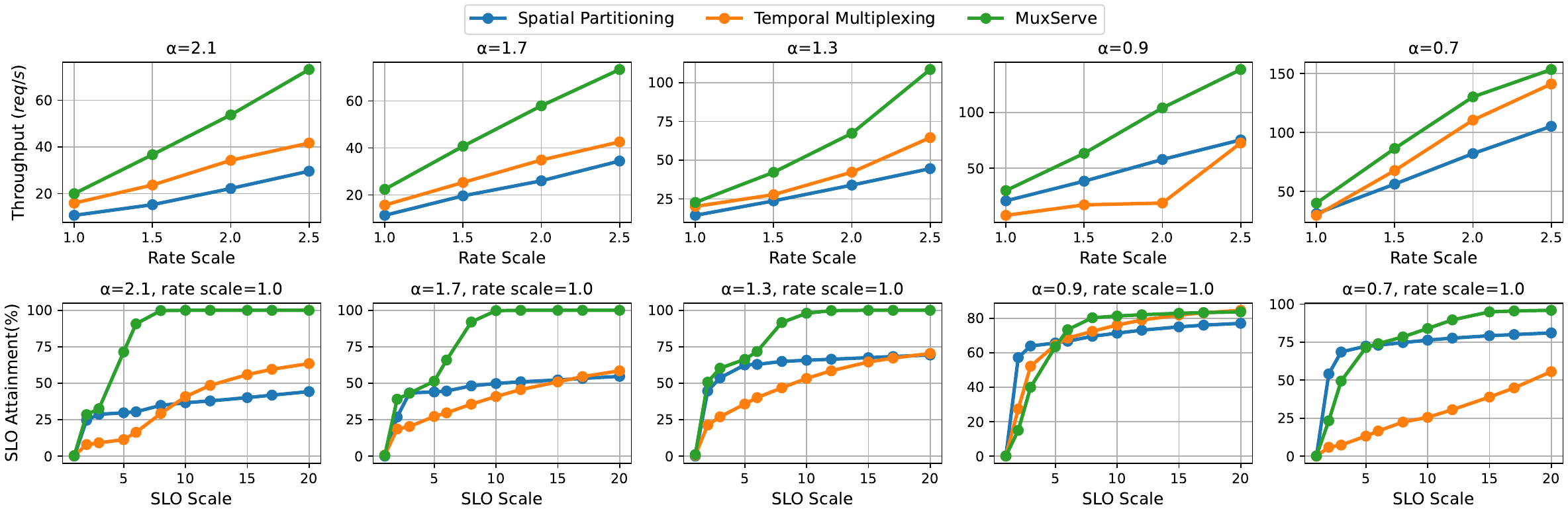}
  \caption{Throughput and SLO attainment on synthetic workloads.}
  \label{fig:e2e_syn}
\end{figure*}

\textbf{Implementation.} 
\Sys is built atop vLLM~\cite{vllm}, an efficient single LLM serving system based on PyTorch~\cite{pytorch}, and utilizes NVIDIA MPS~\cite{mps} to partition SM resources. The major component of \Sys is implemented in a global scheduler, which runs on each LLM unit. The global scheduler manages a request queue and runs the ADBS algorithm to schedule requests from the queue. \Sys disaggregates the prefill and decoding phases and launches separate vLLM processes, configured with different number of SM resources with NVIDIA MPS, as runtime engines for them. The global scheduler schedules prefill or decoding jobs to runtime engine processes via python mutiprocessing shared memory. \Sys additionally runs a memory manager process to manage the unfied memory space. The runtime engines access to the memory space via CUDA IPC handler. When the runtime engine needs to allocate some KV cache for execution, it requests new cache blocks from the memory manager and fills the KV cache to the assigned blocks. After the execution of prefill jobs, the memory manager maintains the allocated KV cache blocks for later usage of corresponding decoding jobs. The KV cache will be released only when the request is finished. We implement the memory manager with C++ to optimize the overhead of block allocation.

\begin{table}[]
\caption{The number of LLMs to be served in different sizes.}
\label{tab:models}
\vskip 0.05in
\centering
\begin{tabular}{c c c c c}
 \hline
 Scale & 4B-8B & 8B-21B & 21B-41B & 41B-70B \\
 \hline
 \#LLMs & 12 & 4 & 2 & 1 \\ 
 \hline
\end{tabular}
\vskip -0.1in
\end{table}

\section{Evaluation}
In this section, we evaluate \Sys on both synthetic and real workloads. We also perform ablation studies to verify the effectiveness of individual components.

\subsection{Experimental Setup}
\textbf{Cluster.} We conduct experiments on a 4 node cluster, each is equipped with 8 NVIDIA A100 (80GB) GPUs. The intra-node connection is 600GB/s NVLink, and the inter-node connection is 200Gbps IB.

\textbf{Metrics.} We use the \textit{aggregated throughput} as our evaluation metric since our target is to evaluate the GPU utilization. Since different LLMs have different arrival rates, we use the rate to compute a weighted average of throughput. 
We also evaluate the \textit{SLO attainment} to study \Sys's effect on latency. SLO attainment measures the percentage of requests that can be finished within the latency target. In this paper, we scale the latency to different multiplies of single device execution latency (i.e. SLO scale).

\textbf{Baselines.} We compare \Sys with two baselines. The first is widely used spatial partitioning, which serves each LLM separately on a group of GPUs. We serve each LLM with vLLM~\cite{vllm}, a state-of-the-art serving framework. Another baseline is temporal multiplexing similar to AlpaServe~\cite{alpaserve}. Since AlpaServe does not support multiplexing of multiple LLMs, we implement this baseline by ourselves with our unified KV cache. For temporal multiplexing, we colocate LLMs with the placement optimized by our placement algorithm, schedule LLMs with first-come-first-serve in a temporal manner, and batch the requests of each LLM with continuous batching.

\begin{figure*}[!htb]
    \begin{minipage}{0.2\textwidth}
     \centering
     \includegraphics[width=\linewidth]{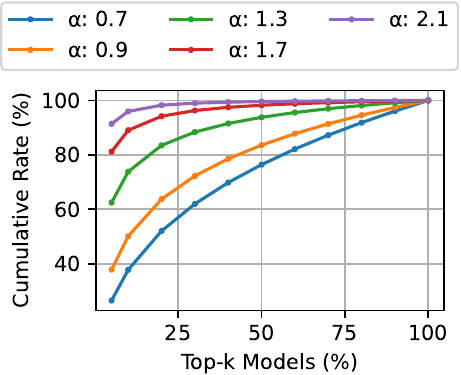}    
     \caption{Cumulative rate distribution as we vary $\alpha$.}
     \label{fig:alpha}
   \end{minipage}\hfill
   \begin{minipage}{0.37\textwidth}
     \centering
     \includegraphics[width=\linewidth]{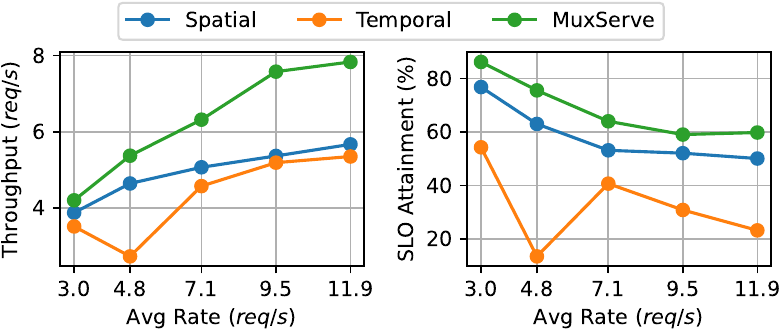}
     \caption{Throughput and SLO attainment as we vary the rates on real workloads.}
     \label{fig:e2e_real}
   \end{minipage}\hfill
   \begin{minipage}{0.37\textwidth}
     \centering
     \includegraphics[width=\linewidth]{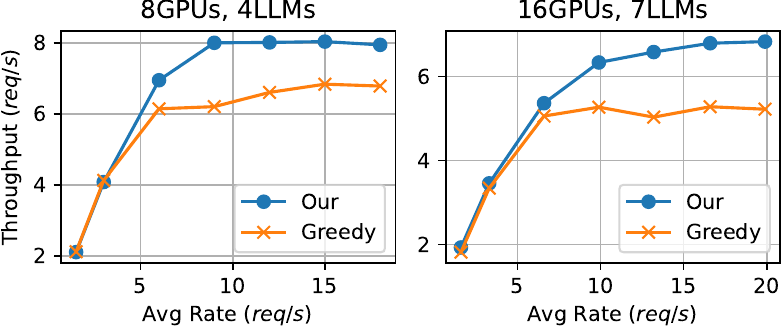}
     \caption{Ablation study of placement algorithm.}
     \label{fig:ab_placemnt}
   \end{minipage}
\end{figure*}

\subsection{End-to-End Results for Synthetic Workloads}

\textbf{Models.} LLaMA~\cite{llama} is the most popular LLM architecture. According to the sizes of LLaMA models, the LLMs can be divided into 4 size buckets. \Cref{tab:models} shows the number of LLMs to be served in different sizes. 

\textbf{Workloads.} For synthetic workloads, we first generate request rates for each LLM using power-law distribution with an exponent $\alpha$, then sample the arrival time of each request with poisson processes. The requests are sampled from ShareGPT. We vary $\alpha$ and rate scales to evaluate a diverse workloads. For each $\alpha$, we first set the maximal request rate for each LLM to $20$ $req/s$, and then scale up the max rate and average rate for evaluation. The $\alpha$ decides the popularity of LLMs, and a larger $\alpha$ means that the fewer LLMs are more popular and receive a higher rates. \Cref{fig:alpha} shows the LLM traffic distribution as we vary $\alpha$. Typically, $\alpha=0.9$ or $2.1$ represent $20\%$ LLMs receives $50\%$ or $90\%$ of the total request rates.




\textbf{Results.} \Cref{fig:e2e_syn} shows the throughput and SLO attainment with varying $\alpha$ and average rates. The throughput of \Sys outperforms two baselines in all scenarios, achieving up to $1.8\times$ improvement. \Sys can also process up to $2.9\times$ more requests within $99\%$ SLO attainment. When $\alpha$ is small, the request rates are more even and \Sys can efficiently colocate \textit{prefill}-\textit{decoding} and \textit{decoding}-\textit{decoding} jobs to improve utilization. But the interference of colocation also brings some overhead, leading to a slightly lower SLO attainment with small SLO scale. With a larger $\alpha$, popular LLMs can be colocated with unpopular LLMs to multiplex memory resources, thus achieving a higher throughput with more SMs and larger KV caches. Popular LLMs can process more requests thus achieving a higher SLO attainment. The results also demonstrate that \Sys can achieve much more significant improvement when the popularity among LLMs is more different (i.e. $\alpha$ is larger).
The detailed results of P99 latency please refer to Appendix~\ref{app:latency}.

\subsection{End-to-End Results for Real Workloads}

To evaluate \Sys on real workloads, we sample LLMs and workloads from ChatLMSYS trace and rescale the rates to evaluate \Sys. ChatLMSYS is a web application that serves multiple LLMs of different scales. In this real workload trace, we serve 16 LLMs with 32 GPUs, and $20\%$ popular LLMs get $50\%$ request traffic. \Cref{fig:e2e_real} shows the throughput and SLO attainment under SLO scale $8$. \Sys achieves up to $1.38\times$ and $1.46\times$ higher throughput compared with spatial partitioning and temporal multiplexing, respectively. As we vary the average rates, \Sys always achieves a higher SLO attainment. When the average rate is $4.8$, several LLMs with medium rates are co-located on a large mesh. Temporal multiplexing cannot efficient multiplex these LLMs thus performing quite bad.

\subsection{Ablation Studies}
In this section, we study the effectiveness of our proposed approaches: placement algorithm in ~\Cref{method:place}, \textit{adaptive batch scheduling} (ADBS) mechanism in~\Cref{method:fairness} and unified resource manager in~\Cref{method:res_manager}.



\begin{figure}[t]
    \centering
    \begin{subfigure}{0.49\textwidth}
	\includegraphics[width=\linewidth]{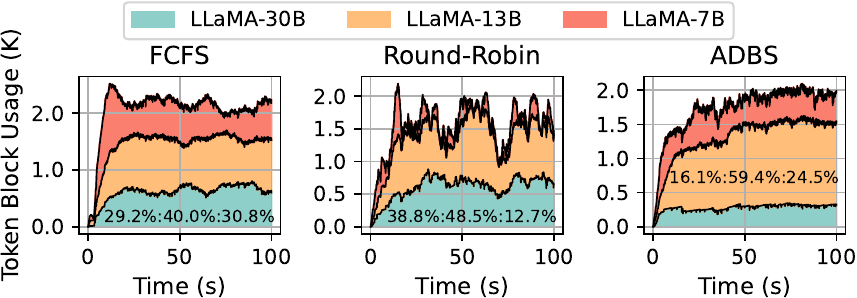}
        \caption{Three LLMs Serving.}
	\label{fig:three_llms}
    \end{subfigure}
    \centering
    \begin{subfigure}{0.49\textwidth}
	\includegraphics[width=\linewidth]{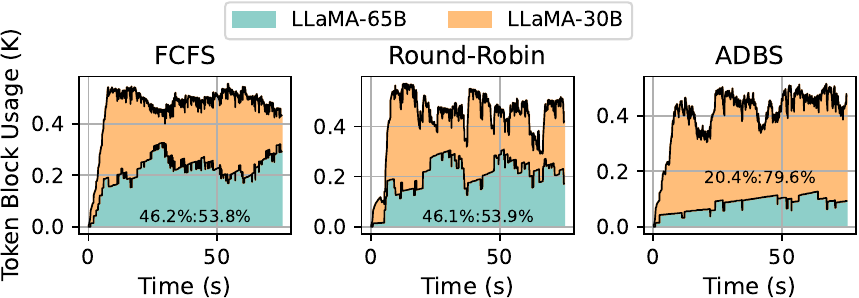}
        \caption{Two LLMs Serving.}
	\label{fig:two_llms}
    \end{subfigure}
  \caption{Comparison of cache usage of different schedule approaches on 4 GPUs. The relative proportion of token block usage is annotated in the figure. FCFS: First Come First Serve, ADBS: ADaptive Batch Size. (a) Request rate: 2:8:8 $req/s$. Throughput ($req/s$): FCFS ($3.8$), Round-Robin ($4.1$), ADBS ($6.2$). (b) Request rate: 1:8 $req/s$. Throughput ($req/s$): FCFS ($3.2$), Round-Robin ($4.9$), ADBS ($6.6$).}
  \label{fig:fairness}
\end{figure}

\textbf{Effectiveness of placement algorithm.} 
To show the effectiveness of our placement algorithm, we conduct a comparison with a greedy algorithm. The greedy algorithm prioritizes the placement of LLM with high arrival rates and greedily assigns it to the mesh with the largest available free memory. The ablation study is conducted on two scales: 8 GPUs with 4 LLMs and 16 GPUs with 7 LLMs. For each scenario, $50\%$ LLMs are popular and occupy more than $70\%$ request traffic. The request arrivals follow poisson distribution. As shown in \Cref{fig:ab_placemnt}, our placement algorithm achieves $1.3\times$ higher throughput compared with greedy algorithm in the right subfigure.

The optimized placement also verifies our insight to prioritize the placement selection for LLMs with large computation requirements. In the 8 GPUs and 4 LLMs case, \Sys colocates two popular small LLMs with one less popular small LLM on 4 GPUs, while allocating 4 GPUs for the less popular large LLM. In the 16 GPUs and 7 LLMs case, \Sys colocates two large LLMs with low arrival rates on 4 GPUs, and splits the rest (majority) of GPUs into groups of 2 or 4 GPUs. \Sys places the other 5 LLMs with high arrival rates on different groups considering their computation demands. To understand this, consider the contrary: if we prioritize the placement selection for LLMs with large memory requirements (i.e. try to balance memory consumption across GPUs). Large LLMs would be prioritized to be placed on large GPU groups without considering their popularity. Then small LLMs will be placed in the manner that tries to fit in the remaining GPU memory space. This is counterintuitive since popular small LLMs could need massive GPUs to fulfill the request traffic requirements.

\textbf{Effectiveness of ADBS.}
We compare \Sys's ADBS to First Come First Serve (FCFS) and Round-Robin in two serving settings to verify the effectiveness (\Cref{fig:fairness}). In \Cref{fig:three_llms}, the average request length is $2:1:1$ for LLaMA-30B, 13B, and 7B, respectively. In \Cref{fig:two_llms}, the average request length is $4:1$ for LLaMA-65B and 30B, respectively. 
In both scenarios, the token block usage of ADBS is more closely aligned with the distribution of arrival rates, thus achieving a \emph{fairer memory resource sharing}. ADBS also achieves $1.43\times$ and $1.85\times$ higher throughput compared with Round-Robin and FCFS scheduling, since the unfair sharing of KV cache blocks hurts the performance of popular LLMs. In addition, FCFS cannot efficiently multiplexing different LLMs.

\begin{figure}[t]
    \centering
    \includegraphics[width=\linewidth]{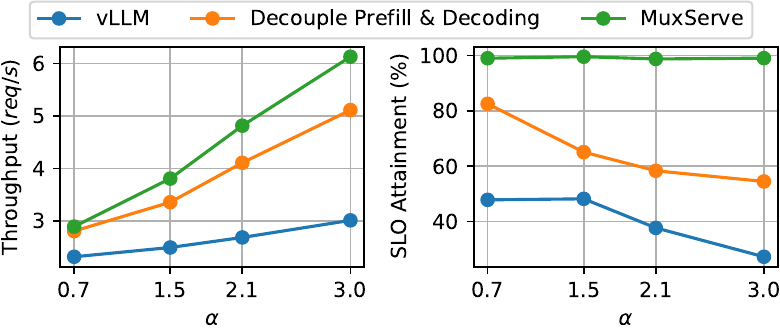}
  \caption{Ablation study of the unified resource manager.}
  \label{fig:urm_eff}
\end{figure}

\textbf{Effectiveness of resource manager.} We study the effectiveness of \Sys's unified resource manager by gradually enabling computation and memory management. We serve 4 LLMs on 4 GPUs and generate arrival rates using power law distribution. \Cref{fig:urm_eff} shows the throughput and SLO attainment (SLO scale $8$) as we vary $\alpha$. By separating prefill and decoding phases with computation resource management, the throughput improves $1.7\times$. With a unified memory manager, \Sys achieves another $1.2\times$ higher throughput and improves SLO attainment by $3.6\times$.

\section{Related Work}

\textbf{DL serving systems.} A wide variety of deep learning serving systems have been proposed to improve the serving efficiency~\cite{tfserving, triton-server}. Recent works~\cite{clipper, nexus, clockwork, shepherd, infaas} utilize temporal multiplexing and introduce better batching and scheduling strategies to improve GPU utilization and meet SLO target. These approaches focus on small DNN models and ignores the parallelism needed in serving LLMs. A more related work is AlpaServe~\cite{alpaserve}, which explores the space of parallelism and serves multiple large DNN models with temporal multiplexing. However, AlpaServe is not designed for LLMs and misses the characteristics of LLM inference phases.


\textbf{LLM serving systems.} Recent years, the outstanding performance of LLMs has aroused strong interests in LLM serving systems. Some prior approaches customize GPU kernels to optimize transformer model inference, for example TensorRT-LLM~\cite{trtllm} and LightSeq~\cite{lightseq}. Recent works such as FasterTransformer~\cite{fastertransformer}, DeepSpeed-Inference~\cite{ds-infer}, vLLM~\cite{vllm} and TGI~\cite{TGI} incorporate intra- and inter-operator parallelism to accelerate LLM inference on multiple GPUs. In addition, memory management~\cite{vllm}, offloading~\cite{flexgen}, iteration-level batching~\cite{orca}, speculative decoding~\cite{specinfer} and cheap instances~\cite{spotserve} have been introduced to enhance the throughput and reduce the cost of LLM serving. \Sys is orthogonal to these works since they focus on optimizing single LLM inference.

\textbf{GPU sharing.} GPU sharing mainly can be categorized into temporal~\cite{zico, wavelet, antman} and spatial sharing~\cite{mig-eval, muxflow, reef}. Salus~\cite{salus} proposes fast job switching and memory management to facilitate temporal sharing. NIVIDIA MIG~\cite{mig} and MPS~\cite{mps} are native support to multiplex jobs on GPUs. GSLICE~\cite{gslice} proposes a dynamic GPU resource allocation and management framework to improve utilization. To overcome the inefficiency of temporal or spatial sharing, Gpulet~\cite{gpulet} introduces mixed spatial-temporal sharing to multiplexing jobs. These works target on multiplexing small DNN jobs, while \Sys explores flexible spatial-temporal multiplexing in emerging LLM serving application.

\section{Conclusion}
In this paper, we introduce \Sys, a flexible and efficient spatial-temporal multiplexing system to serve multiple LLMs concurrently. \Sys colocates LLMs considering their popularity and colocates prefill and decoding jobs leveraging their characteristics to improve GPU utilization. 

\section*{Acknowledgements}
We thank Chang Chen, Qianchao Zhu and Jinming Ma for their insightful review and discussions. The research is supported in part by Shanghai AI Laboratory, CUHK Interdisciplinary AI Research Institute, and the Centre for Perceptual and Interactive Intelligence (CPII) Ltd under the Innovation and Technology Commission (ITC)’s InnoHK. 

\section*{Impact Statement}
This paper presents work whose goal is to advance the field of 
Machine Learning. There are many potential societal consequences 
of our work, none which we feel must be specifically highlighted here.

\bibliography{bibliography}
\bibliographystyle{icml2024}

\newpage
\appendix
\onecolumn
\section{Appendix}

\subsection{P99 Latency Comparison}
\label{app:latency}

\Cref{fig:e2e_lat} shows the P99 average latency, TPOT (Time-Per-Output-Token) and TTFT (Time-To-First-Token) latency of three multiplexing approaches on synthetic workloads. The results demonstrate that \Sys can achieve a lower P99 average latency compared with prior approaches. The P99 TPOT latency of \Sys is a little bit higher than spatial partitioning due to interference, and significantly lower compared wth temporal multiplexing. The P99 TTFT latency of \Sys is lower than both spatial partitioning and temporal multiplexing since \Sys can significantly reduces queuing time. When $\alpha$ is large, the P99 TTFT latency is dominated by the most popular model. Temporal multiplexing can reduces the queuing time, thus obtaining a lower P99 TTFT compared with spatial partitioning.

\begin{figure*}[h]
    \centering
    \includegraphics[width=\linewidth]{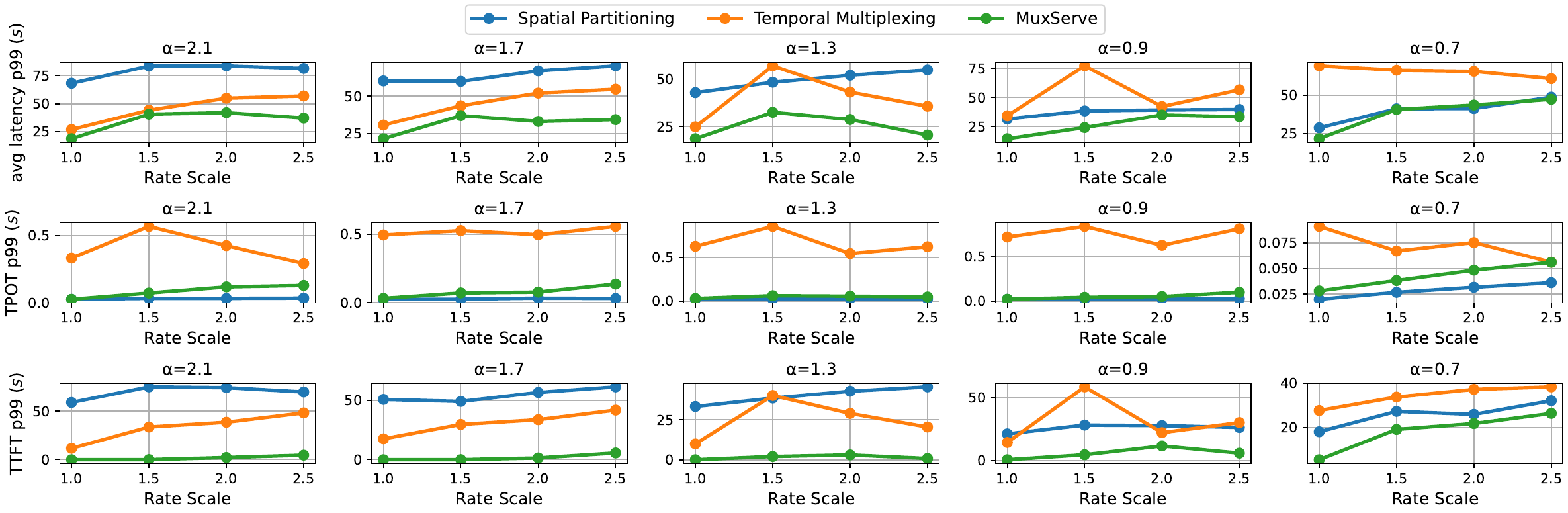}
  \caption{The P99 average latency, TPOT (Time-Per-Output-Token) and TTFT (Time-To-First-Token) latency on synthetic workloads.}
  \label{fig:e2e_lat}
\end{figure*}

\subsection{Throughput Estimator}
\label{app:cost_model}

\Cref{fig:e2e_lat} shows an approximated execution timeline of \Sys, which motivates the formulation of \Cref{eq:tpt}. In a stable serving setting, requests arrive at a fixed interval. The prefill requests are executed one by one, with their decoding steps overlapping. Since the system is stable, it completes some requests every interval and receives new requests. Therefore, we can estimate the execution time of a stable batch $b^{m}$ as $\sum_{i\in b}t^{i}{p} + t^{m}{d}$. Hence we can estimate the throughput with \Cref{eq:tpt}.

\begin{figure}[h]
    \centering
    \includegraphics[width=0.5\linewidth]{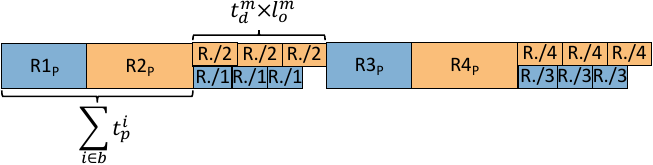}
  \caption{Execution timeline of \Sys in a stable serving setting. $R.$ represents other requests in the batch.}
  \label{fig:e2e_lat}
\end{figure}

\end{document}